\documentclass[hyper]{JHEP3}
\def\bs{\beta^2}
\def\bq{\beta^4}
\def\cs{\cos^2\theta^*}
\def\cq{\cos^4\theta^*}
\def\ss{\sin^2\theta^*}
\def\sq{\sin^4\theta^*}

\newcommand{\definmath}[2] {\def#1{\ifmmode#2\else$#2$\fi}}
\def\slashchar#1{\setbox0=\hbox{$#1$}
     \dimen0=\wd0
     \setbox1=\hbox{/} \dimen1=\wd1
     \ifdim\dimen0>\dimen1
            \rlap{\hbox to \dimen0{\hfil/\hfil}}#1
     \else
            \rlap{\hbox to \dimen1{\hfil$#1$\hfil}}/
     \fi}

\definmath{\etmiss}{\slashchar{E}_T}

\usepackage{epsfig}

\title{Exploring Small Extra Dimensions at the Large Hadron Collider}

\author{B.C.~Allanach$^{\$}$, K.~Odagiri$^\dag$,
M.J.~Palmer$^\ddag$, M.A.~Parker$^\ddag$,
A.~Sabetfakhri$^\ddag$
and B.R.~Webber$^{\ddag}$\\
$^\$$LAPTH, 8
Chemin de Bellevue, B.P. 110, Annecy -le-vieux,  France~74941

\\
$^\dag$Theory Group, KEK, Oho 1--1, Ibaraki 305--0801, Japan\\
$^\ddag$Cavendish Laboratory, University of Cambridge, Madingley Road,
Cambridge, CB3~0HE, UK}

\abstract{
Many models that include small extra space dimensions predict graviton states
which are well separated in mass, and which can be detected as resonances in
collider experiments. It has been shown that the ATLAS detector at the Large
Hadron Collider can identify such narrow states up to a mass of 2080 GeV in the
decay mode $G\rightarrow e^+e^-$, using a conservative model. This work extends
the study of the $e^+e^-$ channel over the full accessible parameter space, and
shows that the reach could extend as high as 3.5 TeV. It then discusses ways in
which the expected universal coupling of the resonance can be confirmed using
other decay modes. In particular, the mode
$G\rightarrow
\gamma\gamma$ is shown to be measurable with good precision, which
would provide powerful confirmation of the graviton hypothesis.
The decays  $G\rightarrow \mu^+\mu^-, W^+W^-, Z^0Z^0$ and jet--jet
are measurable over a more limited range of couplings and masses.
Using information from mass and cross-section measurements, the underlying
parameters can be extracted. In one test model, the size of the
extra dimension can be determined to a precision in
length of $7\times 10^{-33}$ m.
\\
}


\preprint{Cavendish-HEP-02/18\\
CERN-TH/2002-227\\
LAPTH--941/02\\
KEK--TH--851\\
ATL-COM-PHYS-2002-043\\
hep-ph/0211205}

\begin{document}

\section{Introduction}
An exciting idea to be tested in high-energy collider experiments is the
possible existence of narrow graviton resonances in the TeV energy range.
Such resonances are predicted in models with small extra spatial
dimensions. An example is the localized gravity model of Randall and Sundrum
(RS) \cite{randallsundrum}.  This model aroused great theoretical interest
because it motivates the weak-Planck scale hierarchy via an exponentially
suppressed warp factor in a non-factorisable geometry.  Many
possible extensions and elaborations of this type of theory are being
discussed in the
literature~\cite{egs1,egs2,egs3,egs4,egs5,egs6,egs7,egs8,Kogan:2000xc}. Problems
with negative tension brane instability in the original RS model are solved in
some other models. For example, placing the branes on fixed points of orbifolds
projects out the negative energy modes~\cite{rubakov}. An extra scalar with
couplings on the branes can be used to naturally stabilise the brane
separation~\cite{GW}. Warped extra dimensions (with associated graviton
resonances) have also been considered in the context of
supersymmetry~\cite{SUSY-RS1,Gherghetta:2000qt,SUSY-RS2,SUSY-RS3}.  Thus the
discovery of TeV-scale graviton resonances remains a possibility that needs to be
considered seriously in preparing for future collider experiments.

In~\cite{allanach}, the detection of a narrow graviton resonance
using the ATLAS detector at the Large Hadron Collider was considered.  The
main aim of that paper was to establish the discovery limit in the most
favourable decay channel, $G\to e^+e^-$.  A similar study for the CMS detector
has since been reported \cite{Traczyk:2002jh}. In~\cite{allanach} the
angular distribution of the lepton pair was also studied, and it was shown
that the spin-2 nature of the resonance could be confirmed up to a mass
somewhat below the discovery limit.

Apart from its unique spin, the most striking characteristic of the graviton
is its universal coupling to all types of matter and gauge fields.  In the
present paper we consider the accuracy with which the couplings of a narrow
graviton resonance to leptons, electroweak bosons, hadronic jets and Higgs
bosons could be measured at the LHC.  As in~\cite{allanach}, we use the
expected properties of the ATLAS detector as a guide to experimental
limitations and the simplest RS model to characterise the resonance
parameters, but our results should apply to other general-purpose
detectors and to a broad class of models.  We do, however, assume that
all matter and gauge fields are confined to the physical brane and do not
propagate into extra dimensions, thus excluding models of the type considered
in~\cite{Davoudiasl:1999tf,Davoudiasl:2000wi}.

In the simple RS scenario, a 5-dimensional
non-factorizable geometry is used, with two 3-branes of opposite
tension. A graviton Kaluza-Klein spectrum is created, with a scale
\begin{equation}\label{eq:lambda}
\Lambda_\pi=\overline{M}_{Pl} e^{-kr_c\pi}
\end{equation}
where $\overline{M}_{Pl}$
is the reduced effective 4-D Planck scale, $r_c$ is the
compactification radius of the extra dimension, and $k$ is a
scale of the order of the Planck scale. The
geometrical exponential factor (the `warp factor')
generates TeV scales from fundamental Planck scales and hence
motivates the weak-Planck hierarchy, if $kr_c\approx
12$.

The masses of the graviton resonances are given by
\begin{equation}\label{eq:masses}
m_n=kx_n e^{-kr_c\pi}=x_n (k/ \overline{M}_{Pl})\Lambda_\pi
\end{equation}
where $x_n$ are the roots of the Bessel function of order 1
($x_n= 3.8317, 7.0156, 10.1735$ for $n=1,2,3$). The massive graviton
excitations couple with equal strength to the visible sector \cite{Davoudiasl:1999jd}.
However, the higher modes being suppressed by the falling parton distribution
functions, only the lightest mode is considered in this paper.
This does not in any way affect the generality of the approach, as the
analysis can be applied to any such resonances, including the higher modes, so
long as the resonances are narrow and sufficiently separated from the other
modes. This is in contrast to studies in which many excitations, each with
small coupling, contribute to some scattering process~\cite{Arkani-Hamed:1998rs,Antoniadis:1998ig,Accomando:1999sj,Antoniadis:1999bq,Vacavant:2000wz}.
For brevity, we refer to the first massive resonance, with mass $m_G=m_1$,
as ``the graviton".

In the RS model, the couplings of the graviton are given by
$1/\Lambda_\pi$. The graviton mass is determined by
the ratio $k/ \overline{M}_{Pl}$.
Our results are presented in the plane of~$m_G$, $\Lambda_\pi$ to allow
comparisons to be made with any model in this class. We have also defined a
specific test model (identical to that used in \cite{allanach}), with
a value of $k/ \overline{M}_{Pl}=0.01$ (at the bottom of the range suggested in
\cite{Davoudiasl:1999jd}), which according to \cite{Davoudiasl:2000wi} is on the edge of
95\% exclusion for a first graviton excitation mass of
less than 2000 GeV.
Thus, we assign a low coupling constant to
the graviton, and hence obtain a conservative estimate of the
production cross section. This choice leads naturally to a narrow
resonance. This test scenario is used to illustrate the
potential physics reach in each decay channel.
As already emphasised, the results derived do not depend
on the validity of this particular scenario, but can be applied to any model
giving rise to narrow well-spaced graviton resonances.
For example, our results in Section \ref{sec:leptons} show that in a model with
$k/ \overline{M}_{Pl}=0.1$ ($\Lambda_\pi=10$~TeV), the discovery limit in the
$e^+e^-$ channel rises to 3.5 TeV, as a consequence of the increased production
cross section. Values of $k/ \overline{M}_{Pl}>0.1$ are disfavoured on
theoretical grounds because the bulk curvature becomes too large
\cite{Davoudiasl:2000wi}.
%The reach of the LHC is
%presented in the plane of graviton mass versus inverse coupling strength, so
%that the results can be compared with any model containing narrow graviton
%resonances.

An event generator capable of simulating the production and decay
of spin-2 resonances has been developed. This generator is now part
of the standard {\small HERWIG} \cite{HERWIG6,HERWIG6.4} simulation package
(versions 6.2 and later). The generated events are passed
through the ATLAS fast simulation ({\small ATLFAST}
\cite{ATLFAST2}), in order to give a realistic description of
detector resolution and efficiency.

In the following sections, the event generator is described (section
\ref{sec:event}), followed by studies of graviton decays to leptons (section
\ref{sec:leptons}), photons and massive vector bosons (section \ref{sec:IVBs}),
hadronic jets (section \ref{sec:jets}), and Higgs bosons (section
\ref{sec:higgs}). Finally, the ability of the LHC to determine model parameters,
including the length scale of the extra dimension, is discussed.  Here again we
use the simplest RS scenario for illustration.

\section{The event generator}
\label{sec:event}

The implementation of the graviton resonance in the {\small HERWIG} event
generator has been described in~\cite{allanach}. The graviton decays are treated
as $2\to2$ processes, consisting of the two hard production subprocesses
$q\bar q\to G$ and $gg\to G$, followed by the graviton decay.
The relevant matrix elements were computed from the
Feynman rules given in \cite{Giudice:1998ck,Han:1998sg}.
Interference with Standard Model background processes is neglected.
For the range of parameter values considered here, the resonance is
so narrow that its observed width is determined by the detector
resolution in all decay channels, and hence interference effects
cancel in all observable distributions. Note, however, that the
neglect of interference is not a good approximation for the broad
resonances considered in \cite{Traczyk:2002jh}.

The production cross-sections in \cite{allanach} were calculated using the
parton distribution functions (PDFs) of Owens~\cite{Owens:1991ej}, set 1.1.
The present work uses the more recent MRST~\cite{Martin:1998np} PDFs.
This change has no effect on the conclusions of \cite{allanach}.

% The angular distributions of the possible subprocesses, in the centre-of-mass
%frame of the resonance, are shown in Table \ref{grav_table}. Here $\theta^*$
%is the angle between the decay electron and the beam direction in the dilepton
%centre-of-mass frame. The spin-2
%distributions contrast strongly with the angular distributions resulting from
%spin-1 or scalar resonances.
%%
%
The resulting graviton production cross section at the LHC for the test
model is shown as a function of the graviton mass in Figure~\ref{fig:cs}.
The dashed curve shows the predictions for the `central gluon' and
`high gluon' leading-order PDFs of~\cite{Martin:1998np}; the solid
curve is for the average of these, which is the default PDF set for
{\small HERWIG} version 6.3 \cite{Corcella:2001pi}. This has been found to give
the best agreement with recent next-to-leading order fits \cite{Thorne}.
The dashed curves give an indication of the uncertainty due to reasonable
variation of the gluon PDF.
We see that, even with our conservative choice of the coupling, for a
graviton mass of 1.5 TeV the expected number of produced gravitons
is about 5000 for an integrated luminosity of 100 fb$^{-1}$,
falling to about 70 at a mass of 3 TeV.

\FIGURE{
\hbox{\epsfysize=10cm
\epsffile{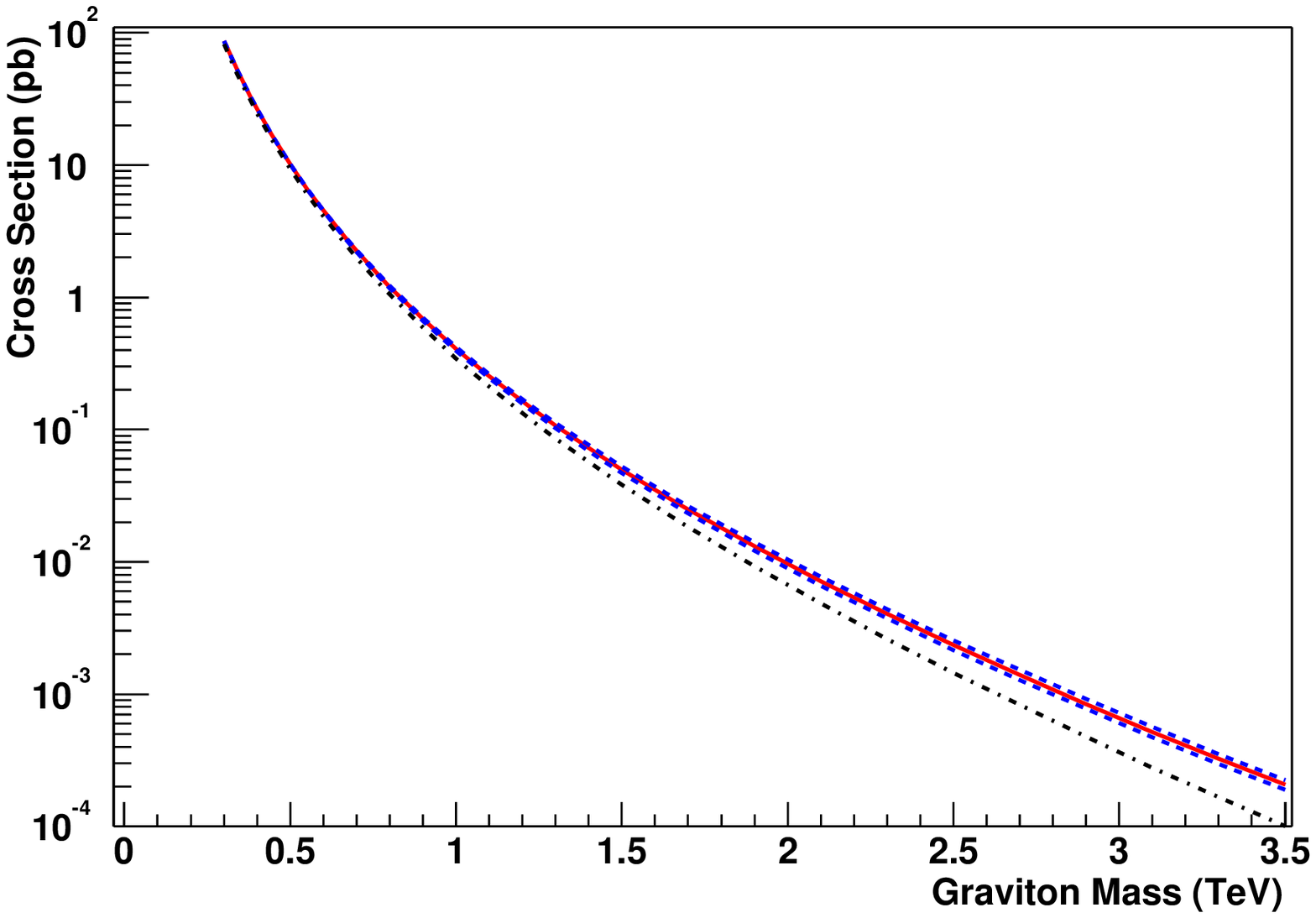}
}
\caption{\label{fig:cs}Cross section for graviton production at LHC.}
}

The dot-dashed curve in Figure~\ref{fig:cs} shows the gluon fusion
contribution to the cross section for the default PDF set; this is shown
as a fraction of the total production cross section in
Figure~\ref{fig:glu_frac}.  Gluon fusion dominates the cross section
for graviton masses up to 3.4 TeV.  This has important implications for
the angular distribution of the graviton decay (see below).
\FIGURE{
\hbox{\epsfysize=10cm
\epsffile{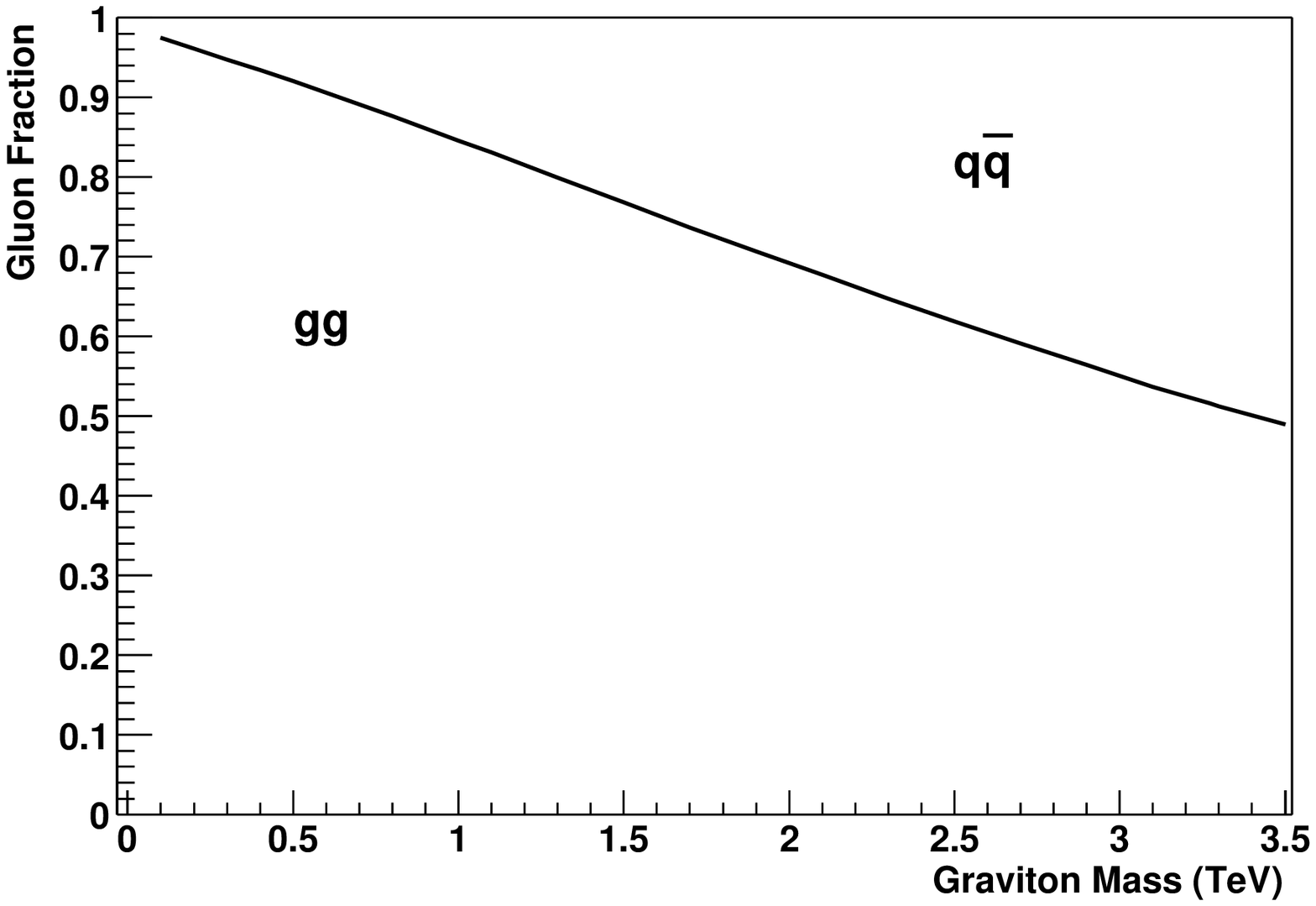}
}
\caption{\label{fig:glu_frac}Contributions of gluon-gluon and quark-antiquark
fusion to graviton production at LHC.}
}

The branching fractions of the graviton into various decay modes
are shown in Figure~\ref{fig:br}.  These predictions are rather
model-independent, depending only on the universality of the coupling. We
see that decays into quark and gluon jets will predominate, due to their high
multiplicity of colour, spin and flavour states.
The Higgs boson fraction depends significantly on the assumed Higgs mass
when $m_G<10 m_H$; we have used $m_H=115$~GeV, the default {\small HERWIG} value.
Out of 5000 produced gravitons with mass 1.5 TeV, we expect roughly 3500
jet-jet, 100
$e^+e^-$, 100 $\mu^+\mu^-$, 100 $\tau^+\tau^-$, 300 $\nu\bar\nu$,
200 $\gamma\gamma$, 450 $W^+W^-$, 225 $Z^0Z^0$ and 15 $H^0H^0$ decays.

\FIGURE{
\hbox{\epsfysize=10cm
\epsffile{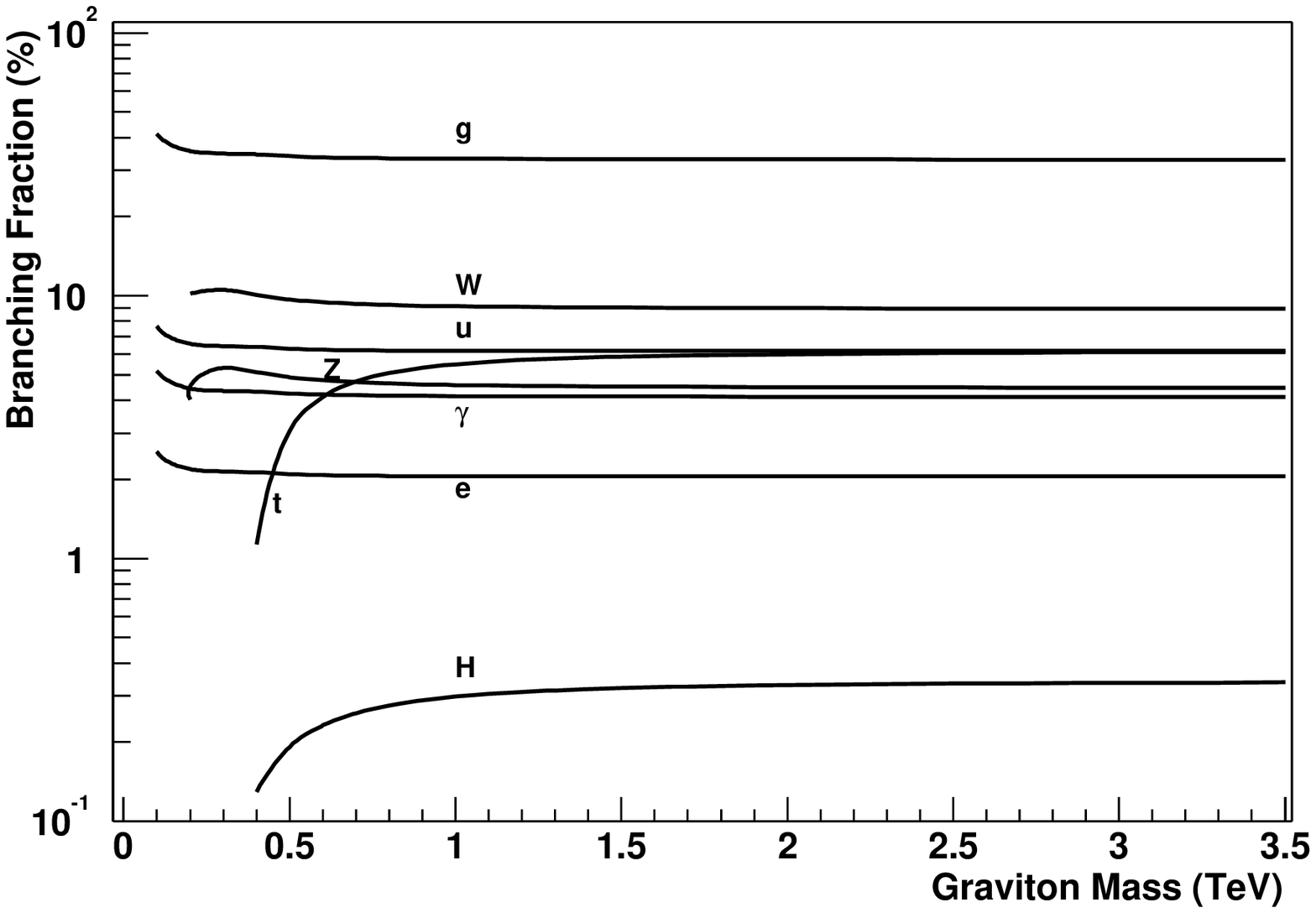}
}
\caption{\label{fig:br}Graviton branching fractions.}
}

The angular distributions of the various decay modes in the graviton rest
frame are summarized in Table~\ref{tab:angdis}. Here $\beta$ represents
the velocity of the decay products, $\beta=\sqrt{1-4m^2/m_G^2}$ for
particles of mass $m$.
In Table~\ref{tab:angdis}, the plot letters refer to Figure~\ref{fig:angdis},
which shows the distributions in the limit of negligible mass ($\beta=1$).  Note
that the angular distribution depends strongly on the production mechanism.
As we saw above, gluon fusion predominates, but the contribution of
quark-antiquark fusion has a structure that tends to flatten the
decay angular distribution.
Notice that the angular distributions of the massive gauge bosons $W$ and
$Z$ are slightly different from those of the massless $\gamma$ and gluon,
even in the limit $\beta\to 1$, owing to their extra longitudinal
polarization state, which has the same distribution as the Higgs
boson.
%The spin-2
%distributions contrast strongly with the angular distributions resulting from
%spin-1 or scalar resonances.

\FIGURE{
\hbox{\epsfysize=10cm
\epsffile{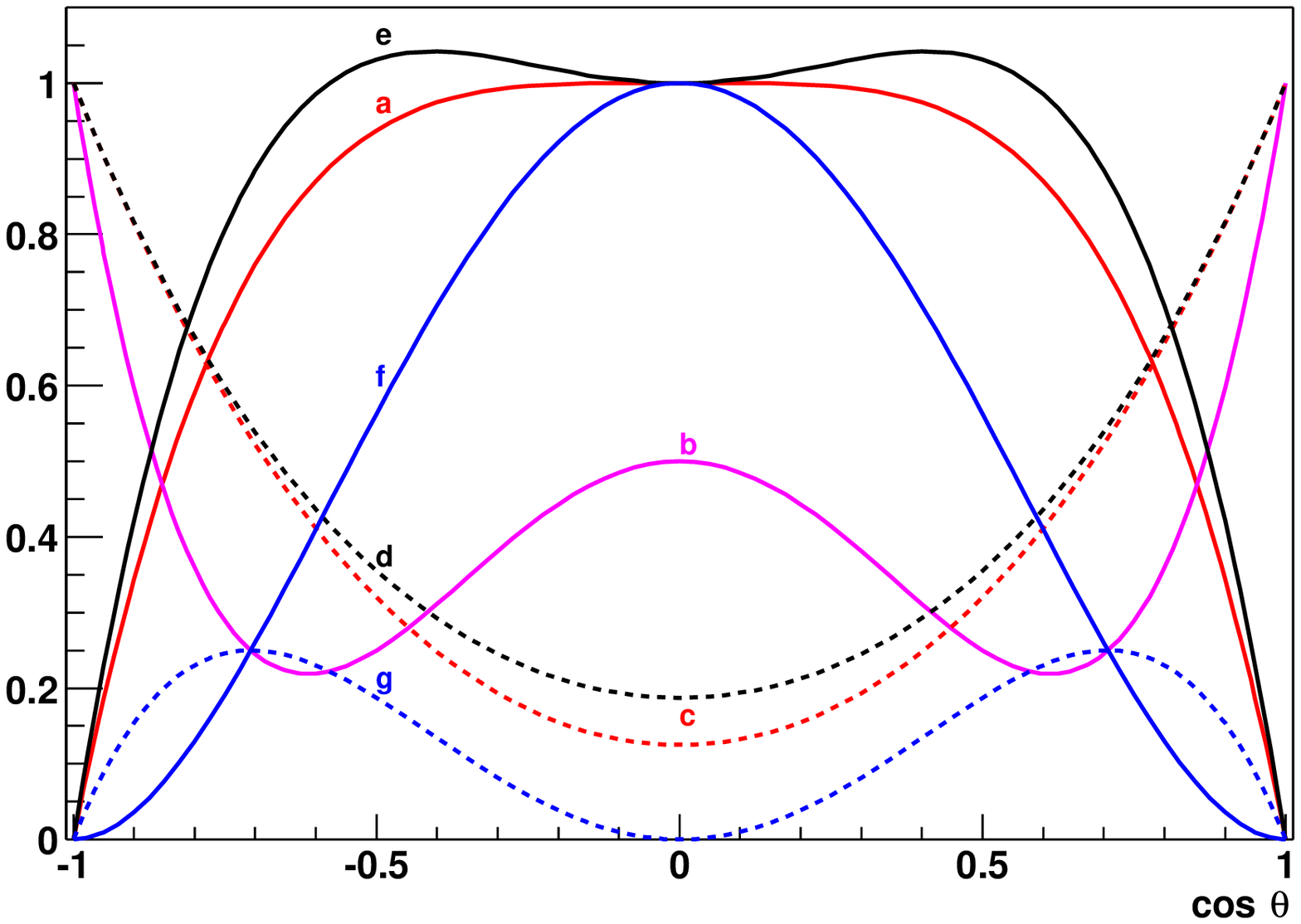}
}
\caption{\label{fig:angdis}Angular distributions in graviton
production and decay ($\beta=1$).}
}

\TABLE{
\renewcommand{\arraystretch}{1.2}
\begin{tabular}{|c|c|c|}
\hline
Process & Distribution & Plot \\
\hline\hline
$gg\to G\to f\bar f$ & $\ss(2-\bs\ss)$ & a \\
$q\bar q\to G\to f\bar f$ & $1+\cs-4\bs\ss\cs$ & b \\
$gg\to G\to \gamma\gamma, gg$ & $1+6\cs+\cq$ & c \\
$q\bar q\to G\to \gamma\gamma, gg$ & $1-\cq$ & a \\
$gg\to G\to WW, ZZ$ & $1-\bs\ss+\frac 3{16}\bq\sq$ & d \\
$q\bar q\to G\to WW, ZZ$ & $2-\bs(1+\cs)+\frac 32\bq\ss\cs$ & e \\
$gg\to G\to HH$ & $\sq$ & f \\
$q\bar q\to G\to HH$ & $\ss\cs$ & g \\
\hline
%$q\bar q$, $gg\rightarrow V\rightarrow f\bar{f}$ & $1+\alpha\cos^2\theta^*$ &
%\\
%$q\bar q$, $gg\rightarrow S\rightarrow f\bar{f}$ & $1$ & \\
%\hline
\end{tabular}
\caption{\label{tab:angdis}Angular distributions in graviton
production and decay. $\theta^*$ is the polar angle of the outgoing fermion in the
graviton rest frame. The letters in the ``plot" column refer to the curves in
Figure
\ref{fig:angdis}.} }

\section{Measurements of the graviton couplings}
In \cite{allanach} it was shown that the graviton resonance can
be detected up to a mass of 2080 GeV in our test model,
using the process $pp\rightarrow G\rightarrow e^+e^-$. The limits are model
independent as long as the graviton couplings are universal and give rise to
narrow resonances, with widths less than the experimental resolution. The
angular distribution of the lepton pair can be used to determine the spin of
the intermediate state. In our test model, the angular distribution favours a
spin-2 hypothesis over a spin-1 hypothesis at 90\% confidence for
graviton masses up to 1720 GeV. In this work, we consider the full range of
parameter space. In some cases, the search reach can be much higher than in the
test model.

In the following, it is assumed that the graviton will be detected in the
$e^+e^-$ channel, with a significance of greater than $5 \sigma$. This study
then assumes that the graviton mass is known, which allows signals with
significances as low as $3\sigma$ to be used in the determination of the
couplings.

\section{Graviton decays to leptons}
\label{sec:leptons}

\subsection{$G\rightarrow e^+e^-$}
This channel offers the best chance of discovery of a graviton resonance at
the LHC, by virtue of the relatively small background from Drell-Yan
processes, and the excellent mass resolution provided by the ATLAS
electromagnetic calorimeter. More details of the proposed discovery search can
be found in \cite{allanach}.

Measurements of the graviton coupling were
studied by simulating signals at graviton masses between 0.5 and 4.0 TeV.
These signals were superimposed on the expected background from Drell-Yan.
Electrons were selected with $p_T>5$~GeV inside the acceptance of the ATLAS
tracking detector ($|\eta|<2.5$), using the standard electron reconstruction
algorithm of {\small ATLFAST}, which accounts for the effect of nearby particles
on the calorimeter signature. The pair with the highest
$p_T$ were used to construct the graviton mass. The mass distribution of the
electron pair is well fitted by a Gaussian signal on a background of the form
$\alpha m_{ee}^{-\beta}$, where
$m_{ee}$ is the mass of the electron pair, and $\alpha$ and $\beta$ are free
parameters. The acceptance of the detector varies from 91 to 76\% across the mass
range, with an estimated systematic error of
$<1\%$, and a negligible statistical error. The efficiency for detecting an
isolated electron is taken as $90\%$. The systematic error on this value will
depend on the details of the detector and reconstruction code used, and is
hence beyond the scope of this study. However, we note that, for the very high
energy electrons involved in these decays, it will not be necessary to make
tight cuts on the electron tracks, and so a high efficiency with a small error
should be obtained. The fit for a graviton mass of 1500 GeV is shown in
Figure~\ref{resonance}.

\FIGURE{
\hbox{\epsfysize=10cm
\epsffile{ee_massfit_1500.epsi}
}
\caption{The number of events per 4 GeV mass bin from a graviton resonance,
with
$m_G=1.5$ TeV (signal), superimposed on the expected Standard Model background
(SM), for 100 fb$^{-1}$ of integrated luminosity. The fit to the data is shown
by the dotted curve.}
\label{resonance}
}

In order to estimate the precision  which could be expected for the
measurement of the production cross-section times branching ratio $\sigma .
B$, a procedure for subtracting the background under the peak is required.
To avoid assumptions about the background shape, we use a simple
background subtraction procedure.  The background estimate $N_B^{est}$ is
obtained by counting the number of events in two bins of width
$w/2$ on either side of the signal. This procedure will work well since
the mass window used to select the signal, of width $w$, is narrow. It
contains
$N$ events, made up of $N_S$ signal and $N_B$ background events.
%Therefore the signal estimate $N_S^{est}$ is given by
%$N-N_B^{est}=N_S+N_B-N_B^{est}$. Since we expect that $N_B^{est}\approx N_B$,
%we can take the statistical errors on $N_B$ and $N_B^{est}$ to be the same.
The error on the signal estimate $N_S^{est}$ is then given by
\begin{equation} \label{eq:error}
\Delta N_S^{est}=\sqrt{\Delta N^2 + \Delta N_B^{est~2}}~.
\end{equation}
%where $\Delta N_S=\sqrt{N_S}$ is the statistical error on the signal and
%$\Delta N_B=\sqrt{N_B}$ is the statistical error on the background estimated
%under the peak from Monte Carlo.
The fractional error on $\sigma . B$ is then
simply equal to
$ \Delta N_S^{est}/N_S^{est} $

In a real experiment, it would be possible to use a more sophisticated
procedure to obtain $N_B^{est}$, by fitting over a larger range of electron
pair mass. This means that our estimates of the
experimental reach are conservative. But since, in the interesting regions,
the background levels are small, the effect of background subtraction on the
final error is also small.

A systematic error arises
from the errors on the acceptance, electron identification efficiency and
luminosity. The method for luminosity measurement in ATLAS is not yet decided,
but an error of 5-10\% is within reach of conventional methods
\cite{ATLAS_TDR}. Methods to improve this to 1-2\% are under consideration.
Together with the as yet unknown electron reconstruction efficiency (see
above) this means that the systematic error on cross-section measurements is
very uncertain at present, and hence we plot our results using statistical
errors only. The effect of the systematic error on the extraction of model
parameters is discussed in section
\ref{parameters}.

The above procedure was used to determine how well the ATLAS detector could
measure the graviton coupling. The {\small HERWIG}/{\small ATLFAST} simulation was run at each
graviton mass, and for the standard model background, for 100 fb$^{-1}$ of
integrated luminosity, corresponding to one year of running of the LHC at its
nominal luminosity of 10$^{34}$ cm$^{-2}$s$^{-1}$. The estimated statistical
errors on $\sigma .B$ are plotted on Figure \ref{ee_contours}. Also shown are
the region excluded by Tevatron data \cite{Abe:1997gt,Abbott:1998rr} and lines of
constant
$k/\overline{M}_{Pl}$. In the test model, with
$k/\overline{M}_{Pl}=0.01$, a 10\% measurement of the production rate is possible
for graviton resonance masses as high as 1400 GeV. 
In models with $k/\overline{M}_{Pl}=0.1$ ($\Lambda_\pi=10$~TeV), a 20\%
measurement of the coupling is possible for graviton masses as high as 3.5 TeV,
indicating the ultimate search reach. 
For completeness, we have continued the contours into the
region of very small
$k/\overline{M}_{Pl}$. This region is excluded in the RS--I model
\cite{Davoudiasl:2000wi}, but may be relevant to other models with low mass graviton
resonances. In such a case, values of $\Lambda_\pi$ as high as 100 TeV are in
principle accessible. For high values of $k/\overline{M}_{Pl} > 0.05$,
horizontally-striped on the plot, the graviton resonance width becomes larger
than the experimental mass resolution, making a measurement of the width
possible. It is unlikely that such a measurement would be possible in any of the
other channels considered here, since their mass resolutions are far inferior.
Once the width becomes very large, it would be necessary to increase the size of
the mass window in order to maintain a high efficiency for the signal.
Interference with the Drell-Yan background would also need to be taken into
account. However, the production cross-section is proportional to
$(k/\overline{M}_{Pl})^2$, and so the detection of the resonance is
trivial in this region.

\FIGURE{
\hbox{\epsfysize=10cm
%\epsffile{ee_contours.epsi}
\epsffile{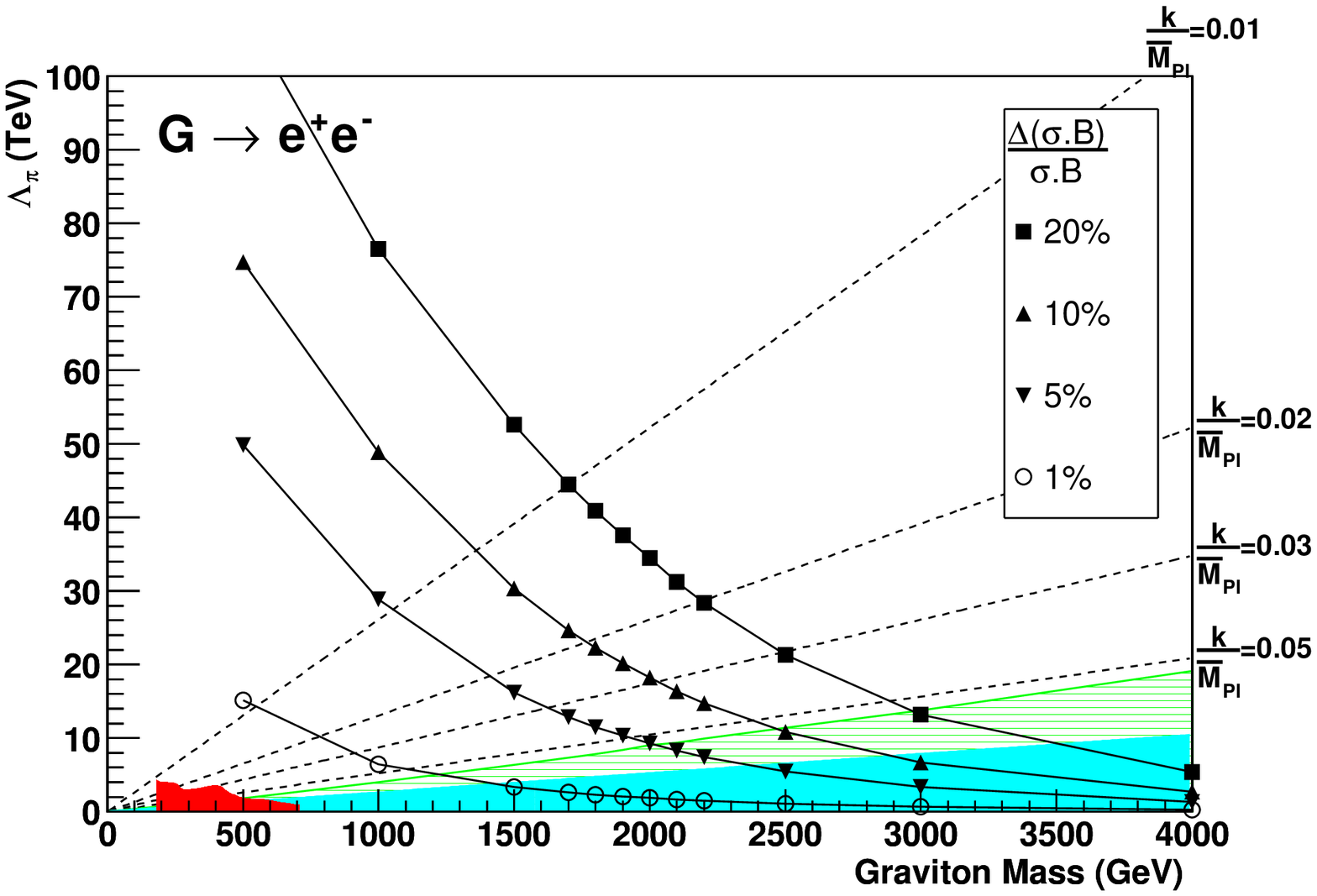}
}
\caption{Contours showing the statistical precision expected for a measurement
of
$\sigma .B$ in the decay mode $G \to e^+e^-$ for 100 fb$^{-1}$ of integrated
luminosity (Solid lines with markers). Also shown are lines of equal
$k/\overline{M}_{Pl}$ (dashed). The test model has $k/\overline{M}_{Pl}=0.01$. The
(red) blocked region near the origin is excluded by Tevatron data
\cite{Abe:1997gt,Abbott:1998rr}. In the (green) horizontally-striped
region above
$k/\overline{M}_{Pl} \approx 0.05$ the resonance width is larger than the
experimental resolution. The (blue)
blocked region with $k/\overline{M}_{Pl}
> 0.1$ is disfavoured theoretically.}
\label{ee_contours}
}

\subsection{$G\rightarrow \mu^+\mu^-$}

The analysis of this channel is very similar to that for the electron case. 
Muons were selected with
$p_T>5$~GeV using the standard muon reconstruction algorithm of {\small ATLFAST}.
The pair with the highest $p_T$ were used to construct the graviton mass. The
discovery potential in the
$\mu^+\mu^-$ channel is not as great as in the
$e^+e^-$ case, because the momentum resolution of the magnetic spectrometer
decreases at high muon momentum. This is reflected in a much poorer mass
resolution for high muon pair masses. In addition, the reconstruction
efficiency is poorer than for electrons.
%For example, for a mass of 1500 GeV, the
%muon pair mass resolution is 89 GeV, compared to 8 GeV for electron pairs.
%This means that far more background lies under the mass peak in the muon case.
%In addition, the reconstruction efficiency for muons is estimated in
%\cite{ATLAS_TDR} to be $85\%$ at high transverse momentum.
Nonetheless, precisions of
$\sigma .B <10\%$ are achievable in the test model case, for graviton resonance
masses up to 1250 GeV. Figure
\ref{mumu_contours} shows the results, which can provide a valuable check of
lepton universality in the graviton couplings. 

\FIGURE{
\hbox{\epsfysize=10cm
%\epsffile{mumu_contours.epsi}
\epsffile{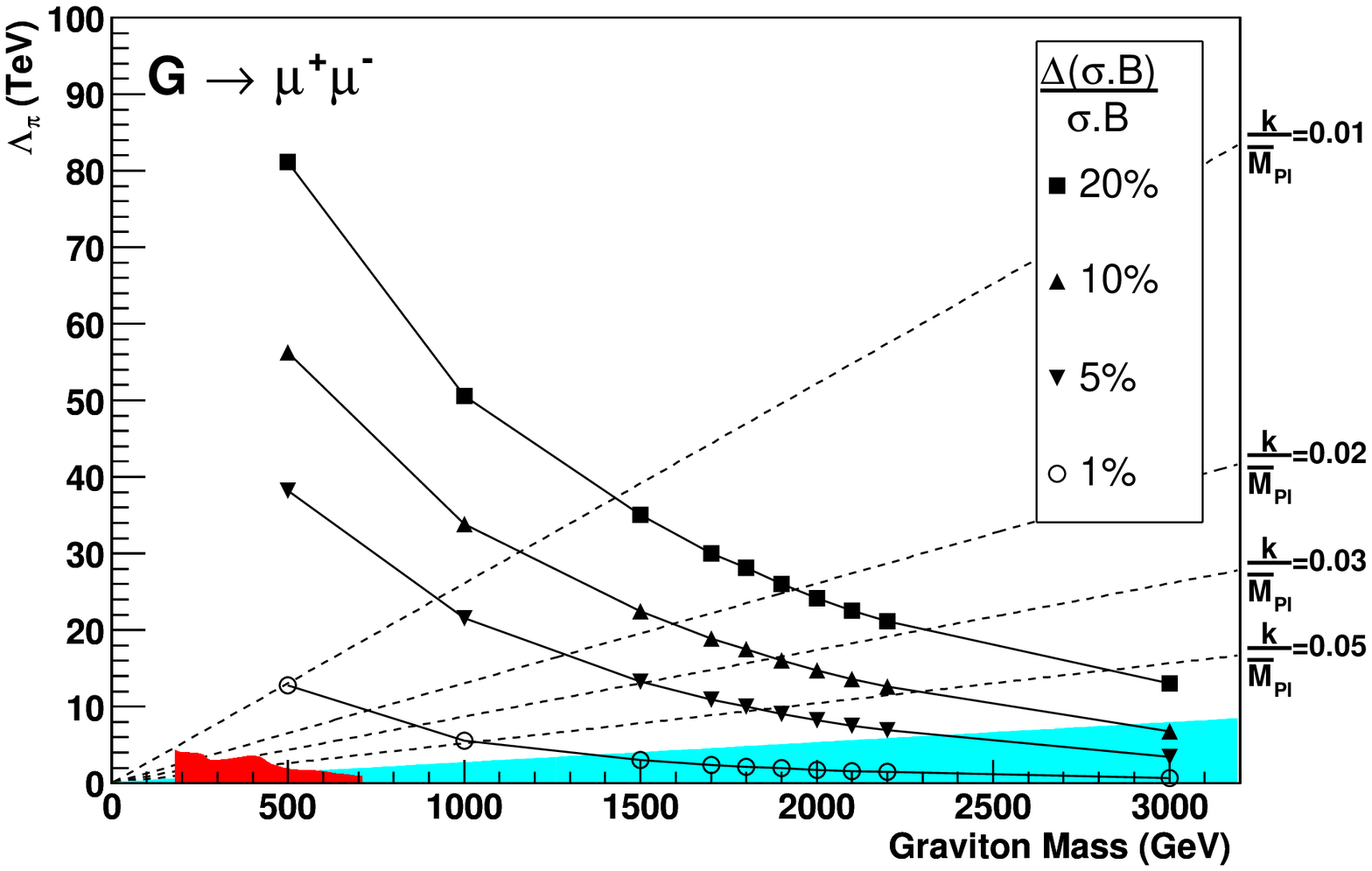}
}
\caption{Contours showing the statistical precision expected for a measurement
of
$\sigma .B$ in the decay mode $G \to \mu^+\mu^-$ for 100 fb$^{-1}$ of
integrated luminosity, as for Figure \ref{ee_contours}.
%(Solid lines with markers). 
%Also shown are lines of equal
%$k/\overline{M}_{Pl}$ (dashed). The test model has $k/\overline{M}_{Pl}=0.01$.
%The blocked region near the origin is excluded by Tevatron data
%\cite{Abe:1997fd,Abazov:2001qd}. In the cross-hatched region at low
%$\Lambda_\pi$ the resonance width is larger than the experimental resolution. 
}
\label{mumu_contours}
}

\subsection{$G\rightarrow \tau^+\tau^-$}
The $\tau^+\tau^-$ decay mode would be extremely hard to observe on the large
standard model background from QCD jets. The
missing energy from the $\tau$ decays would spoil the mass resolution for the
graviton signal, further degrading the significance of any peak. This signal
is therefore not considered further.

\section{Decays to vector bosons}
\label{sec:IVBs}
The detection of decays to vector bosons would be very important in
establishing the nature of a graviton resonance, since the universal coupling
would be very different from that expected for other exotic objects, such as a
$Z^\prime$. In addition the angular distribution of the decay products is a
characteristic signature of the resonance spin (see Table \ref{tab:angdis} and
Figure \ref{fig:angdis}).

\subsection{$G\rightarrow \gamma\gamma$}

The method used to study this channel is identical to that used for the electron
case. The minimum photon $p_T$ was set to 1 GeV. A photon detection efficiency of
$90\%$ was applied after the effect of the standard photon selection cuts in
{\small ATLFAST}.  

The mass resolution for photon pairs is excellent, being very
close to that for electrons. However the background is much less well
understood. The {\small HERWIG} calculation of the cross-section only includes the Born
term and production from gluon-gluon interactions via a quark box diagram. It
is known that these diagrams alone predict a cross-section for photon pair
production at the Tevatron which is a factor of $\approx 5$ too small
\cite{TeVphotons}. This large discrepancy means that one cannot rely on
existing Monte Carlo simulations to produce a reliable background estimate,
and in particular that the angular distribution of the background cannot be
trusted. We note that these effects were not considered in \cite{ref:sridhar}.

As discussed in Section 2, graviton production at accessible masses is
dominated by gluon-gluon fusion, and therefore the photon angular distribution
is strongly forward-backward peaked, as shown by curve c in Figure
\ref{fig:angdis}. An important background, not presently simulated by {\small HERWIG}
(or any other current event generator), is bremsstrahlung from initial
state partons, which is also strongly forward-backward peaked. For these
reasons, we do not attempt an analysis of how well the resonance spin
could be determined at the LHC. However, it should be noted that the
background level and angular distribution can easily be measured in the
experiment by using data away from the resonance itself. The photon
channel will then be very significant in establishing the nature of the
resonance.

For this analysis, we use a photon pair background 5 times as large as
predicted by {\small HERWIG} in order to estimate the precision which could be reached
on the coupling in this channel. 
The results are shown in Figure
\ref{gamgam_contours}.

\FIGURE{
\hbox{\epsfysize=10cm
\epsffile{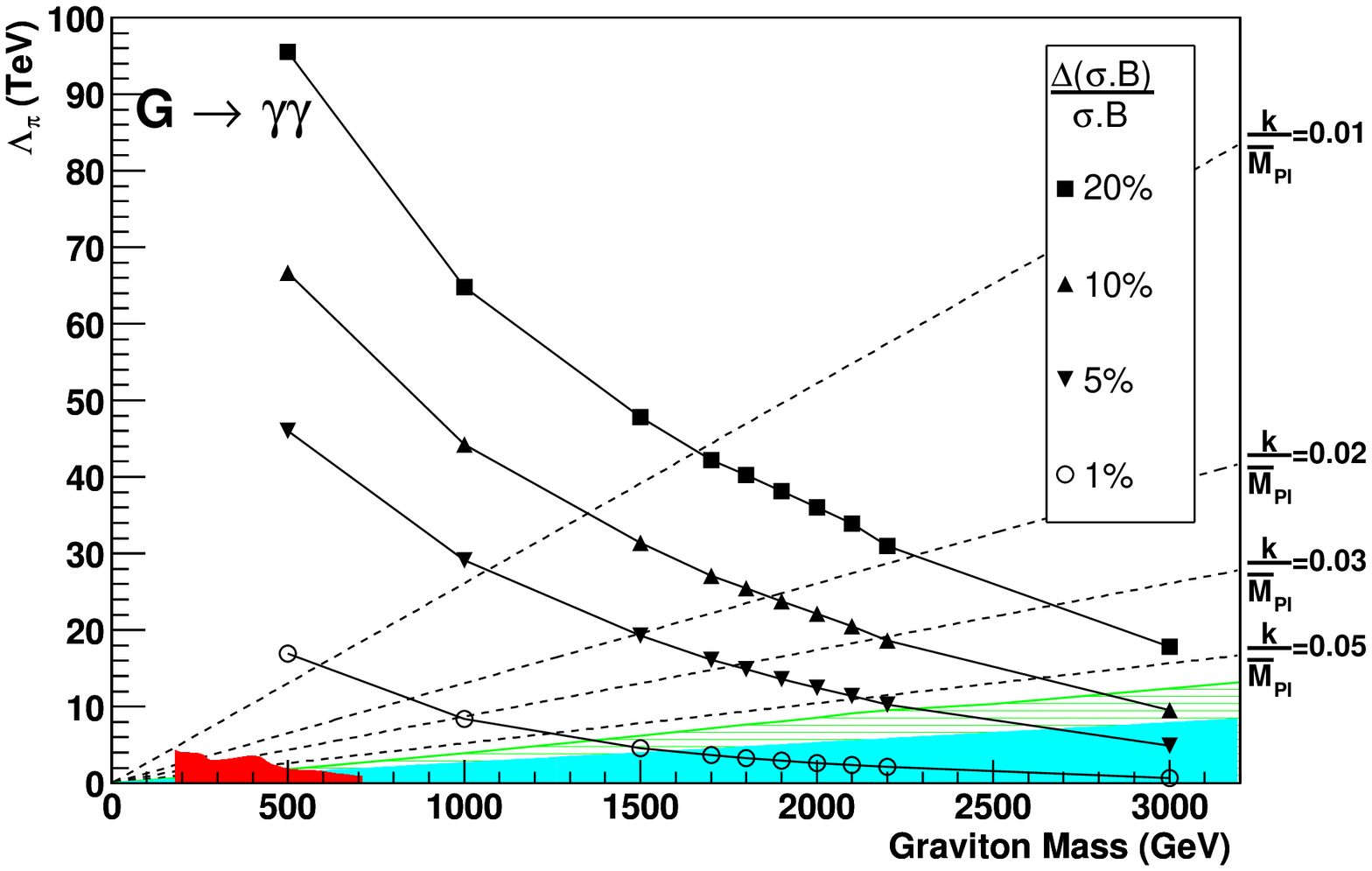}
}
\caption{Contours showing the statistical precision expected for a measurement
of $\sigma .B$ in the decay mode $G \to \gamma\gamma$ for 100 fb$^{-1}$ of
integrated luminosity, as for Figure \ref{ee_contours}. }
\label{gamgam_contours}
}

\subsection{$G\rightarrow WW\rightarrow l\nu jj$}

Graviton decays to a pair of W bosons are best detected in the mode where one
W decays leptonically to the electron or muon final states, and one
hadronically. This mode has a reasonable branching fraction (29\% of W
pairs), and the reconstruction of the events can be performed by assuming
that the missing energy is due only to the neutrino from the leptonic
decay and using the W mass as a constraint.

This channel has considerable difficulties compared to the leptonic
and $\gamma \gamma $ channels due to the large background from $t\bar{t}$
and $W$+2 jets. Consequently this channel would not be a discovery
channel, but would be useful in confirming the universality of the
graviton's coupling. Scale factors for the background were obtained
by comparing the {\small HERWIG} cross-section with the NLL cross-section for
$t\bar{t}$~\cite{Bonciani:1998vc} and by using the prescription in \cite{Giele:1990vh}
for $W$+2 jets, giving scale factors of 1.8 and 1.7 respectively.
Background from $WW$ production was also included for completeness
but was negligible compared to the other backgrounds.

Standard {\small ATLFAST} settings for high luminosity were used except for
the jet reconstruction algorithm, for which the Mulguisin
algorithm~\cite{Mulguisin} with a minimum distance of $\Delta R=0.2$ was
used. This algorithm was chosen because it was found to give the highest
signal reconstruction efficiency.

Jet reconstruction is problematic because the jets come from a $W$
which is highly boosted. This naturally
leads to two jets very close in $\eta -\phi $: for a 3 TeV graviton,
both jets often fall within the same calorimeter cell as
defined by {\small ATLFAST}.

The neutrino 4-momentum is reconstructed from \etmiss\ 
and fixing the mass of the $l\nu $ system to be the mass of a $W$.
This gives a quadratic equation which is solved and the average of the two
solutions taken. The second $W$ is reconstructed from the highest
$p_{T}$ jet and a jet which gives $65<M_{jj}<85$ GeV. If more than one
possibility is found, the combination that gives the highest $p_{T}(W)$ is chosen.

The $p_{T}$ cuts are set fairly low to allow the whole range of
graviton masses above 500 GeV to be treated in a single analysis. Some
improvement could be made by tuning these cuts for a smaller mass region,
using the mass of the resonance as measured in the $e^+e^-$ channel as a
guide.

To reduce the $t\overline{t}$ background the following cuts were
imposed: a top
reconstruction veto which attempts to reconstruct a top mass from
each of the $W$s and an acceptably close jet (within $\Delta R<1)$, and a cut on
the number of central ($|\eta |<2$) jets.

The signal reconstruction efficiency is 22\% at 1.5~TeV, dropping
to 6\% at 3~TeV. The mass resolution is 6\%. The background is not
well described by any simple form over the entire range, but it is
very smooth. Therefore it is expected that background subtraction
will be successfully achieved by fitting a function in the sidebands.
The error on the background estimate was found by fitting an exponential with a window
of $m\pm 2\sigma $ in a fit region $m\pm 6\sigma $.
Figure \ref{fig:WWbgfit} shows the background subtraction procedure, in the case
$k/\overline{M}_{Pl}=0.05$.

\FIGURE{
\hbox{\epsfysize=10cm
\epsffile{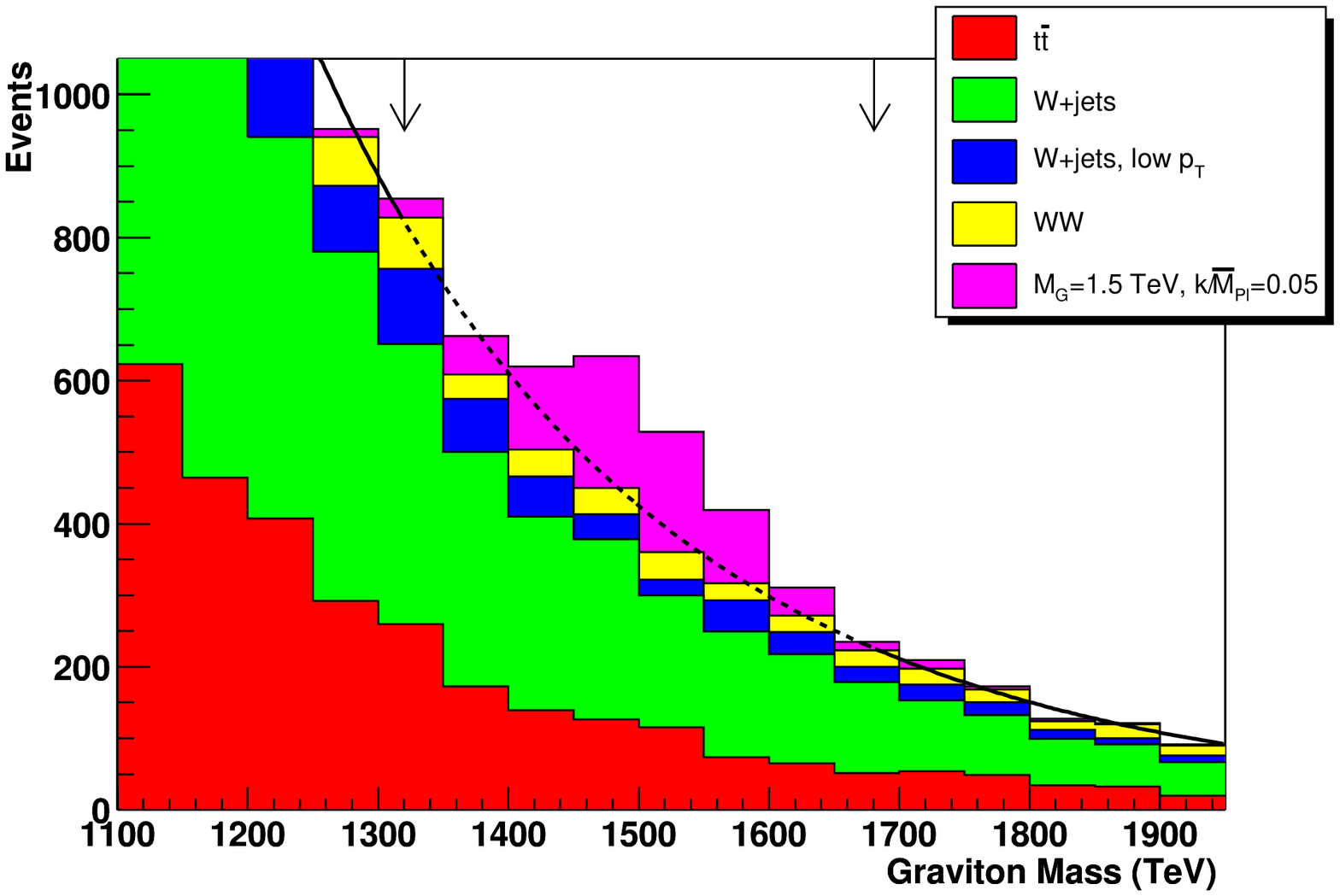}
}
\caption{\label{fig:WWbgfit}
A fit to the background in the sidebands around the $G\rightarrow WW$ signal
(solid line) and its extrapolation under the signal peak (dashed line).} }

The error on the signal estimate, $N_{S}^{est}$
is given by equation (\ref{eq:error}),
%$\Delta N_{B}^{est}$ was estimated using:
%
%\[
%\frac{\Delta N_{B}^{est}}{N_{B}^{est}}=\frac{\Delta
%N_{Fit}}{N_{Fit}}=\frac{1}{\sqrt{N_{Fit}}}\] where $N_{Fit}$ is
taking the error on $N_{B}^{est}$ to be the statistical error on
the number of
events in the fit region.
The estimated statistical errors for $\sigma .B$ were then calculated
in the same way as the other channels and are shown in Figure
\ref{fig:ContoursWW}.

In the test model with $k/\overline{M}_{Pl}=0.01$, a measurement of $\sigma.B$ is
not possible. However, for higher values of $k/\overline{M}_{Pl}$, measurement is
possible with statistical accuracies in the range 5-30\%.
%of up to 5\% for $m_{G}<1.8$~TeV. 
For high
masses, the overlap of jets means that the efficiency of the selection cuts
becomes very low. A different analysis would be required, perhaps based on
event shapes in order to extend the reach further. At low masses, the effects
of the
$p_{T}$ cuts on the determination of the fit parameters can be seen.

\FIGURE{
\hbox{\epsfysize=10cm
\epsffile{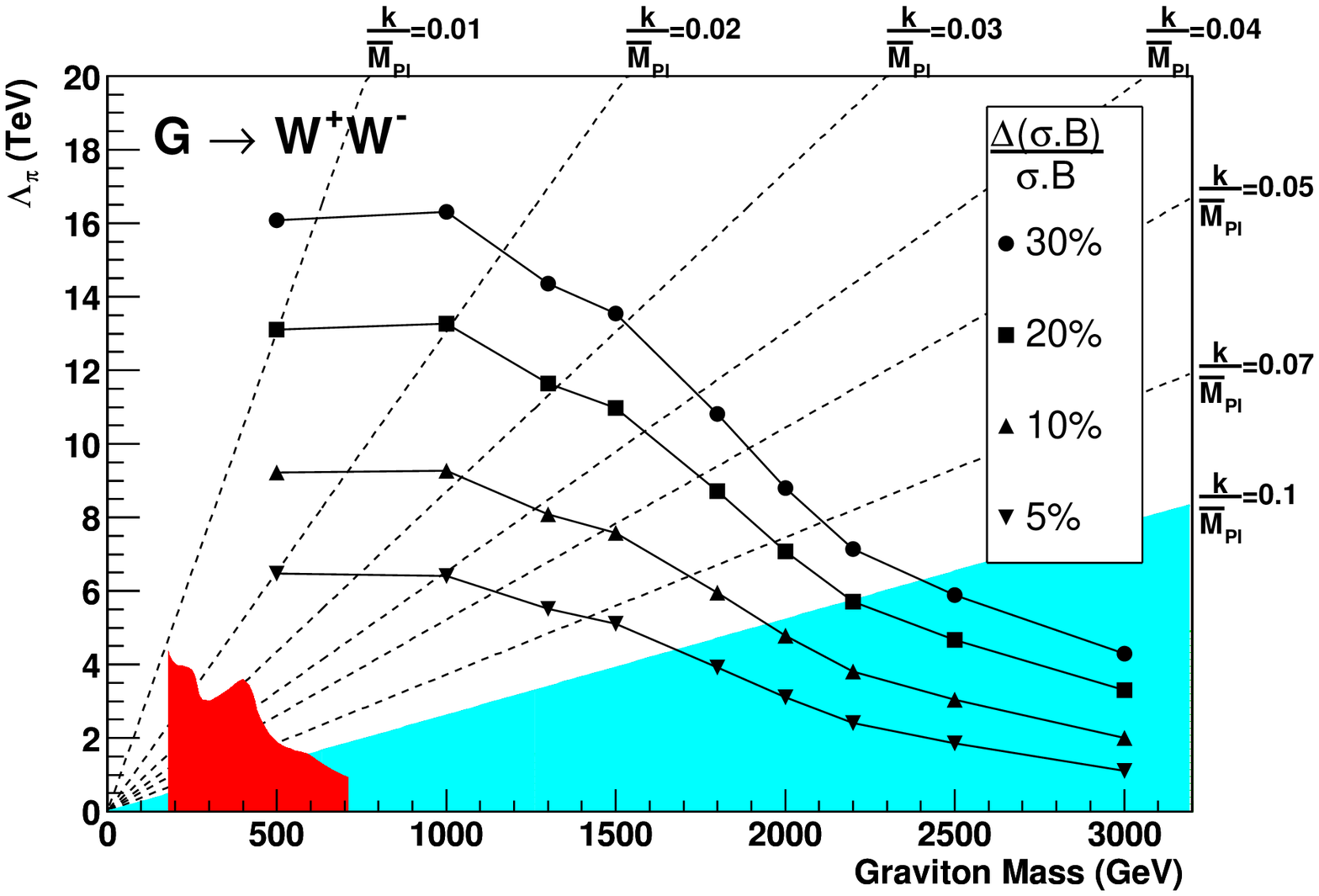}
}
\caption{\label{fig:ContoursWW}Contours of statistical
precision for $\sigma .B$ in the $G\rightarrow WW$
channel, as for Figure \ref{ee_contours}.}
}

\subsection{$G\rightarrow ZZ\rightarrow lljj$}

This channel is analyzed in much the same way as the $WW$ channel.
The principal difference is that the dominant background is $Z$+2 jets
-- there is no $t\bar{t}$ equivalent. Consequently, similar reconstruction
and cuts were employed with the omission of the cuts intended to reduce
the $t\bar{t}$ background.

The same background fitting and subtraction procedure as in the $WW$
case is followed leading to a similar error determination, the results
of which are shown in Figure \ref{fig:ContoursZZ}.

Signal reconstruction efficiency is similar to the $WW$ case and
mass resolution is better, $\sim 3$\%. The smaller branching ratio for this
decay is offset by the lower background and the final reach is very
similar to the $WW$ case. As before, the effect of decreasing efficiency
at high mass is evident as is the effect of the $p_{T}$ cuts at lower
masses.

\FIGURE{
\hbox{\epsfysize=10cm
\epsffile{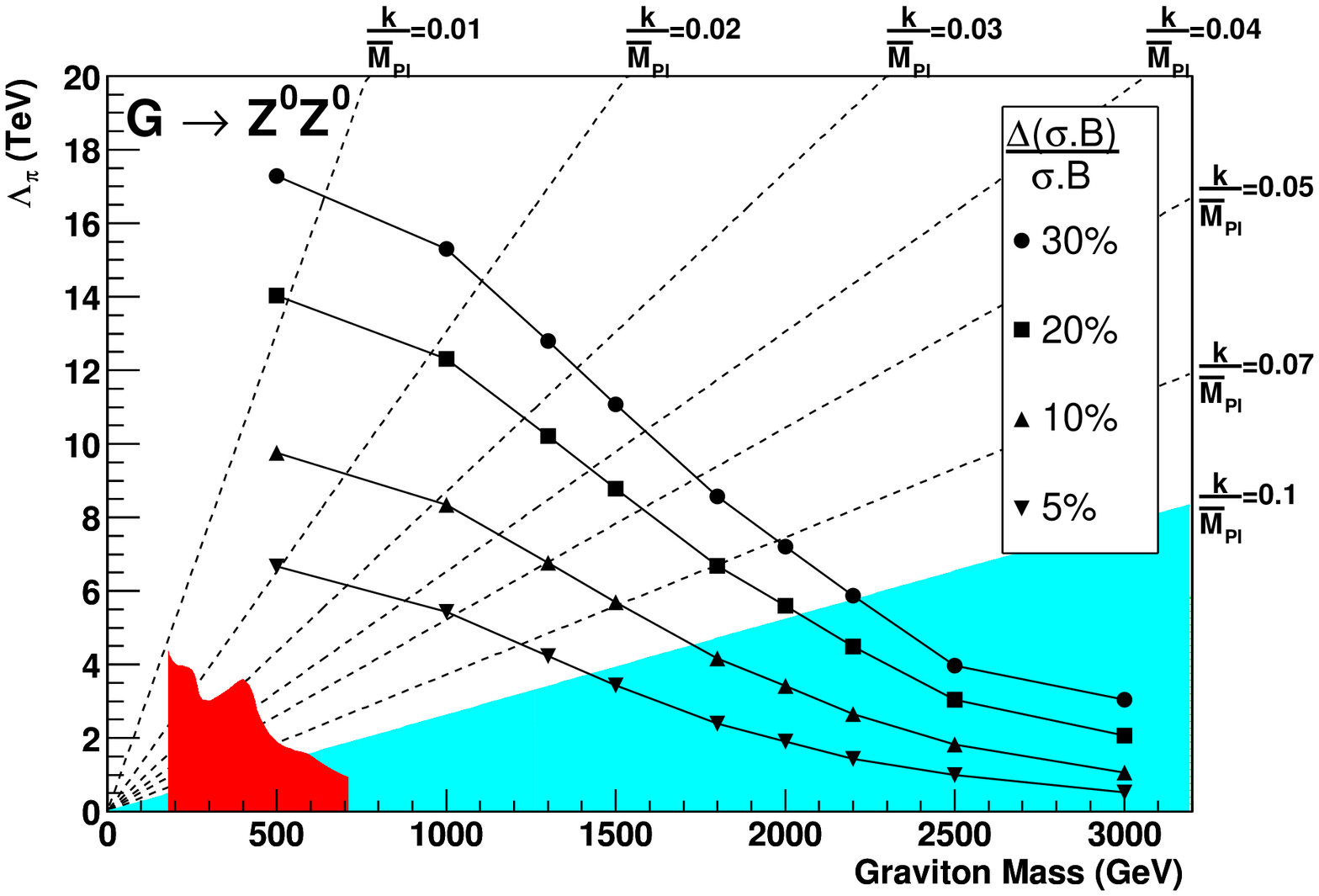}
}
\caption{\label{fig:ContoursZZ}Contours of statistical
precision for $\sigma .B$ in the $G\rightarrow ZZ$ channel, as for Figure
\ref{ee_contours}.} }

\section{Decays to hadronic final states}
\label{sec:jets}
\subsection{Inclusive decays to 2 jets}
The signature for the decay mode $G\rightarrow jj$ comprises two
energetic jets in the detector, producing a large
transverse hadronic energy. The dominant background process to this
topology is QCD multi-jet production, which forms a continuum irreducible 
background. This large background would make it impossible to find a
statistically significant signal peak without the knowledge of the peak
position provided by other channels such as the $e^+e^-$ final state, except
in cases with very large graviton couplings. However, as stated above, we assume
that any discovery will be made in other channels, and investigate the potential
to measure the coupling strength alone in hadronic channels.

\FIGURE{
\hbox{\epsfysize=10cm
\epsffile{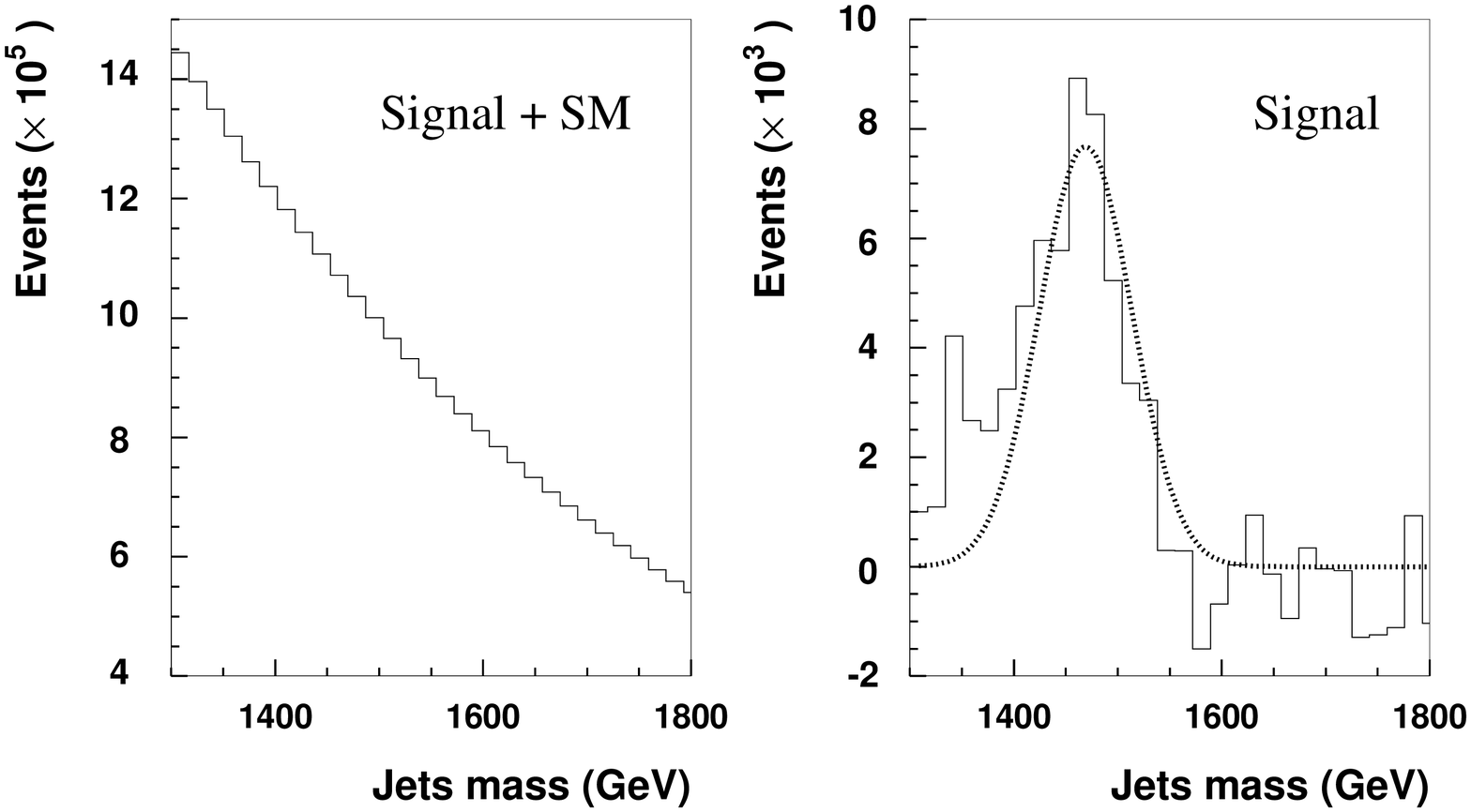}
}
\caption{Distribution of  invariant mass a)
             for all two jet events selected, and b) after background
subtraction.}
\label{Gjjsignal1500}
}

$G\rightarrow jj$ candidates are selected by requiring at
least two jets with minimum transverse energies $(E_T)$ of a quarter of
the graviton mass.
% above 100 GeV. The threshold is raised to 500 GeV for high mass
% gravitons ($M_{G} \ge$ 1.5 TeV).
The continuum background is shown in Figure \ref{Gjjsignal1500}a, with a
resonance at 1500 GeV superimposed, using $k/\overline{M}_{Pl}$ = 0.08.
The signal is not visible to the eye.
The overall acceptance for the signal selection
cuts ranges from
$40\%$ to $60\%$ depending on the model parameters $m_G$ and $\Lambda_\pi$.
% $33\%$ to $80\%$.
Although the signal observability
is not sufficient for a discovery in this decay channel, it is
adequate for measurement of the graviton coupling over much of parameter space;
$N_S/\sqrt{N_B}$ varies  from 3.8 (for $m_{G}$ = 500 GeV, $k/\overline{M}_{Pl}$ =
0.02) to 0.1  (for $m_{G}$ = 2200 GeV, $k/\overline{M}_{Pl}$ =  0.1). $N_S$ and
$N_B$ are the number of $G\rightarrow jj$ signal and
background events after the selection cuts.

 The mass of the graviton is determined from the invariant mass of the
two highest $E_T$ jets; if a third jet is in close proximity ($\Delta\eta,
\Delta\phi <1$) to one of these two high $E_T$ jets, the mass of the graviton is
calculated from the invariant mass of the three highest $E_T$ jets.
The mass resolution for this decay mode degrades dramatically with 
a long tail toward lower masses ($e.g.$, the mass resolution is about 
160 GeV for a 2000 GeV graviton). 
% Therefore, a procedure, independent of the signal shape, has been 
% adapted to extract the signal. 
%An example of a signal from $G\rightarrow jj$
%with a mass of 2000 GeV is shown in figure \ref{Gjj2000_signal}.

To extract the signal, a procedure independent of the signal shape
has been adopted.
% A procedure, independent of the signal shape, has been adapted to 
% extract the signal.
 The mass distribution of the jets is fitted by a Gaussian signal (whose 
peak is fixed to be at the previously determined peak of the signal)
on a background of
exponential form. The  procedure for subtracting the background under the
peak is the same as that for the $G\rightarrow WW$ channel. 
Figure \ref{Gjjsignal1500}b shows the signal after background subtraction.
The very high statistics in the background sidebands allow the background
under the signal to be estimated with very high precision, revealing the
signal peak.
Figure \ref{fig:contour_jj} shows the fractional error on
$\sigma.B$ versus
$\Lambda_{\pi}$ and graviton mass in the $G\rightarrow jj$ decay mode for
100~fb$^{-1}$ of integrated luminosity.

%\FIGURE{
%\hbox{\epsfysize=10cm
%\epsffile{Gjj2000_signal.ps}
%}
%\caption{Distribution of  invariant mass
%             for signal events from the process $G\rightarrow jj$ with
%$m_G =2000$ GeV, showing the skewed mass resolution after jet reconstruction.}
%\label{Gjj2000_signal}
%}

\FIGURE{
\hbox{\epsfysize=10cm
\epsffile{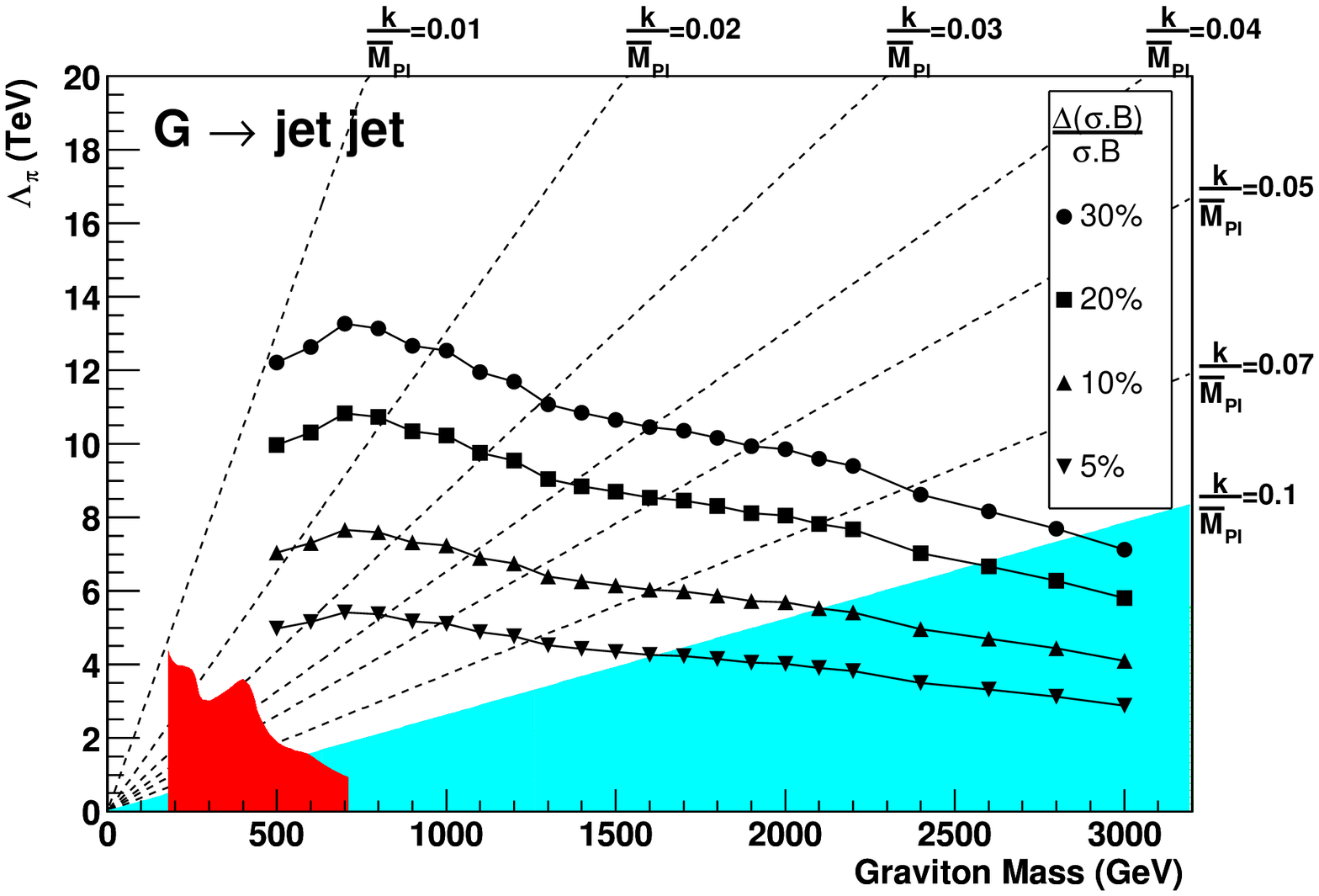}
}
\caption{The fractional error on $\sigma.B$ versus $\Lambda_{\pi}$
             and graviton mass in $G\rightarrow jj$ decay mode
             for 100 fb$^{-1}$ of integrated luminosity, as for Figure
\ref{ee_contours}.}
\label{fig:contour_jj}
}

\subsection{Decays to heavy quarks}
We have considered whether tagging heavy quarks ($t$ or $b$) might be useful to
improve the signal to background ratio and allow a better measurement of the
graviton coupling to quarks. However, since both the graviton and gluon
couplings to quarks are flavour independent, no extra discrimination is
obtained. The best experimental strategy is therefore to exploit the higher
statistical power of the inclusive dijet channel.

\section{Decays to Higgs pairs}
\label{sec:higgs}
In the case that $m_G>2m_H$ then the channel $G\rightarrow HH$ is open. The
angular distribution is given in Table \ref{tab:angdis}. The branching ratio is
1/12 of that into di-photons, if $m_G \gg 2m_H$. The dominant final state will be
4 $b$-quarks (for a light Higgs boson) or 4 $W$'s (for a heavy Higgs boson). Both
of these final states are difficult to reconstruct, and will have poor
statistics. The Higgs coupling is therefore unlikely to be measurable at the LHC.

\section{Summary of the reach for coupling measurements}
The results of the simulations of each graviton decay channel are summarised in
Table \ref{tab:summary}. The precision which can be reached on $\sigma.B$ in each
of the channels investigated, is presented for a range of points in the $m_G,
\Lambda_\pi$ plane.

\TABLE{
\renewcommand{\arraystretch}{1.2}
\begin{tabular}{|c||c|c|c||c|c|c||c|c|}
\hline
     & \multicolumn{8}{c|}{Point $m_G,\Lambda_\pi$ (TeV)} \\
\cline{2-9}
Channel        &1,10  &1,20&1,30&2,10&2,20&2,30&3,10&3,20 \\
\hline\hline
$e^+e^-$       &  1.6 &  3.3 &  5.3 &  5.4 & 11.0 & 17.1 & 15.1 & 30.7 \\
$\mu^+\mu^-$   &  1.9 &  4.5 &  8.2 &  6.2 & 15.2 & 28.2 & 15.1 & 32.7 \\
$\gamma\gamma$ &  1.2 &  2.9 &  5.2 &  3.9 &  8.8 & 15.2 & 10.5 & 23.0 \\
$WW$           & 11.6 & 44.9 &  -   & 38.2 &   -  &   -  &  -   &   -   \\
$ZZ$           & 13.7 & 50.1 &  -   & 52.7 &   -  &   -  &  -   &   -   \\
$jj$           & 19.0 & 77.0 &  -   & 31.0 &   -  &   -  & 59.0 &   -   \\
\hline
\end{tabular}
\caption{\label{tab:summary} The relative precision achievable (in \%) for
measurements of
$\sigma.B$ in each of the channels considered, for fixed points in the
$m_G,\Lambda_\pi$ plane. Points with errors above 100\% are not shown.}
}

\section{Determination of the model parameters}
\label{parameters}

Several models have been built~\cite{egs8,Kogan:2000xc} in which our analysis applies.
They are based on the original RS model but with additional branes.
Supersymmetric versions~\cite{Gherghetta:2000qt,SUSY-RS2} have also been constructed,
in which the graviton resonances are identical to the ones studied here.
These models all have a narrow relatively strongly coupled first massive
graviton mode, and our analysis should apply to them (as long as
the parameters are such that the graviton is narrower than the experimental
resolution).
This is because graviton modes couple to matter in
proportionality to the energy-momentum tensor as a model-independent feature,
guaranteeing the universality of the coupling. The overall coupling strength
is a model dependent parameter, as of course is the connection between the
model parameters and the mass or coupling of the first graviton mode.
We note that even in the factorisable extra dimension case~\cite{Arkani-Hamed:1998rs,Antoniadis:1998ig}, the
resonances would be well separated if the extra dimension were small enough.
The splitting between the resonances in the factorisable case is constant,
being $2 \pi / R$
where $R$ measures the size of the extra dimension(s)~\cite{Giudice:1998ck}.
If $1/R$ were in the range $M_Z-1$ TeV, resonance graviton production would still
not be possible because each state couples with negligible strength, suppressed
by the Planck mass. But if models were constructed which increase this coupling
in the factorisable dimension scenario, our analysis would be fully applicable to
this case also.

As a specific example, we have considered the precision which could be
obtained on the parameters of the RS model, in the case $m_G=1500$~GeV. Since
the model only contains two parameters ($m_G$ and $\Lambda_\pi=39$~TeV), only
two measurements are required to fully constrain the model. 
$m_G$ can be measured directly in the $e^+e^-$ channel, with a
statistical precision of better than 1 GeV (our fit gives 0.7 GeV for
$m_G=1500$~GeV). The energy scale error is given in \cite{ATLAS_TDR} as $<0.7\%$
in this energy range, giving a resolution on $m_G$ of 10.5~GeV. The statistical
error on the coupling depends on
$\Lambda_\pi$, ranging from 1\% at low $\Lambda_\pi$ to 15\% for
$\Lambda_\pi$=39~TeV as in the test model.  The dominant systematic error on
the coupling is due to the luminosity measurement, since the systematic errors on
efficiencies and acceptances will be below the 1\% level. We assume,
conservatively, that the luminosity can be obtained to 10\%, giving an overall
error on $\sigma.B$ of 18\%. We can then infer a value for the compactification
radius of the extra dimension,
$r_c$, and its error, using  equations
(\ref{eq:lambda}) and
(\ref{eq:masses}). The precision on the coupling measurement is then directly
reflected in the error in
$r_c$ giving $r_c= (82
\pm 7) \times 10^{-33}$ m. This reach to extremely small distance scales is a
consequence of the warp factor in the model, working on the TeV-scale
measurements of physical observables.

\section{Conclusions}

The LHC detectors will be capable of discovering narrow graviton resonances
predicted in a range of models with extra space dimensions. Such resonances
will most easily be detected in the di-electron and di-photon final states.
The coupling strength of the resonance to $\mu^+\mu^-, W^+W^-, Z^0Z^0$ and jet--jet final
states (but not $\tau^+\tau^-$ or $H^0H^0$) can also be measured over a wide range of
parameter space.  The resonance spin can also be measured over a more limited
mass range. Taken together, these measurements would provide compelling evidence
for the existence of a massive graviton resonance coupling to the SM fields with a
universal coupling strength.

Since models with a small number of Planck-scale extra dimensions are highly
constrained, the model parameters can be extracted with good precision, if a
particular scenario is assumed. In the RS test model, the size of the extra
dimension can be inferred to better than 10\%, corresponding to a precision in
length of
$7\times 10^{-33}$ m, using measurements of $\sigma.B$ and the graviton mass.

\acknowledgments
We would like to thank I.~Vernon for helpful discussions on different
scenarios. MAP and AS would like to thank C.G.~Lester for his technical help. MJP
wishes to thank the Cambridge e-Science Centre for its assistence in
using the UK e-Science Grid to generate the background samples used in
sections 5.2 and 5.3.
This work was funded by the U.K. Particle Physics and Astronomy Research Council.

\bibliography{Graviton2_v10}

\providecommand{\href}[2]{#2}\begingroup\raggedright\begin{thebibliography}{10}

\bibitem{randallsundrum}
L.~Randall and R.~Sundrum, {\it A large mass hierarchy from a small extra
  dimension},  {\em Phys. Rev. Lett.} {\bf 83} (1999) 3370--3373,
  [\href{http://xxx.lanl.gov/abs/http://arXiv.org/abs/hep-ph/9905221}{{\tt
  http://arXiv.org/abs/hep-ph/9905221}}].

\bibitem{egs1}
L.~Randall and R.~Sundrum, {\it An alternative to compactification},  {\em
  Phys. Rev. Lett.} {\bf 83} (1999) 4690--4693,
  [\href{http://xxx.lanl.gov/abs/http://arXiv.org/abs/hep-th/9906064}{{\tt
  http://arXiv.org/abs/hep-th/9906064}}].

\bibitem{egs2}
C.~Csaki and Y.~Shirman, {\it Brane junctions in the {Randall-Sundrum}
  scenario},  {\em Phys. Rev.} {\bf D61} (2000) 024008,
  [\href{http://xxx.lanl.gov/abs/http://arXiv.org/abs/hep-th/9908186}{{\tt
  http://arXiv.org/abs/hep-th/9908186}}].

\bibitem{egs3}
J.~Lykken and L.~Randall, {\it The shape of gravity},  {\em JHEP} {\bf 06}
  (2000) 014,
  [\href{http://xxx.lanl.gov/abs/http://arXiv.org/abs/hep-th/9908076}{{\tt
  http://arXiv.org/abs/hep-th/9908076}}].

\bibitem{egs4}
I.~Oda, {\it Mass hierarchy from many domain walls},  {\em Phys. Lett.} {\bf
  B480} (2000) 305--311,
  [\href{http://xxx.lanl.gov/abs/http://arXiv.org/abs/hep-th/9908104}{{\tt
  http://arXiv.org/abs/hep-th/9908104}}].

\bibitem{egs5}
I.~Oda, {\it Mass hierarchy and trapping of gravity},  {\em Phys. Lett.} {\bf
  B472} (2000) 59--66,
  [\href{http://xxx.lanl.gov/abs/http://arXiv.org/abs/hep-th/9909048}{{\tt
  http://arXiv.org/abs/hep-th/9909048}}].

\bibitem{egs6}
T.-j. Li, {\it Non-compact {AdS(5)} universe with parallel positive tension
  3-branes},  {\em Phys. Lett.} {\bf B478} (2000) 307,
  [\href{http://xxx.lanl.gov/abs/http://arXiv.org/abs/hep-th/9911234}{{\tt
  http://arXiv.org/abs/hep-th/9911234}}].

\bibitem{egs7}
N.~Arkani-Hamed, S.~Dimopoulos, G.~R. Dvali, and N.~Kaloper, {\it Infinitely
  large new dimensions},  {\em Phys. Rev. Lett.} {\bf 84} (2000) 586--589,
  [\href{http://xxx.lanl.gov/abs/http://arXiv.org/abs/hep-th/9907209}{{\tt
  http://arXiv.org/abs/hep-th/9907209}}].

\bibitem{egs8}
I.~I. Kogan, S.~Mouslopoulos, A.~Papazoglou, G.~G. Ross, and J.~Santiago, {\it
  A three three-brane universe: New phenomenology for the new millennium?},
  {\em Nucl. Phys.} {\bf B584} (2000) 313--328,
  [\href{http://xxx.lanl.gov/abs/http://arXiv.org/abs/hep-ph/9912552}{{\tt
  http://arXiv.org/abs/hep-ph/9912552}}].

\bibitem{Kogan:2000xc}
I.~I. Kogan, S.~Mouslopoulos, A.~Papazoglou, and G.~G. Ross, {\it Multi-brane
  worlds and modification of gravity at large scales},  {\em Nucl. Phys.} {\bf
  B595} (2001) 225--249,
  [\href{http://xxx.lanl.gov/abs/http://arXiv.org/abs/hep-th/0006030}{{\tt
  http://arXiv.org/abs/hep-th/0006030}}].

\bibitem{rubakov}
V.~A. Rubakov, {\it Large and infinite extra dimensions: An introduction},
  {\em Phys. Usp.} {\bf 44} (2001) 871--893,
  [\href{http://xxx.lanl.gov/abs/http://arXiv.org/abs/hep-ph/0104152}{{\tt
  http://arXiv.org/abs/hep-ph/0104152}}].

\bibitem{GW}
W.~D. Goldberger and M.~B. Wise, {\it Modulus stabilization with bulk fields},
  {\em Phys. Rev. Lett.} {\bf 83} (1999) 4922--4925,
  [\href{http://xxx.lanl.gov/abs/http://arXiv.org/abs/hep-ph/9907447}{{\tt
  http://arXiv.org/abs/hep-ph/9907447}}].

\bibitem{SUSY-RS1}
R.~Altendorfer, J.~Bagger, and D.~Nemeschansky, {\it Supersymmetric
  {Randall-Sundrum} scenario},  {\em Phys. Rev.} {\bf D63} (2001) 125025,
  [\href{http://xxx.lanl.gov/abs/http://arXiv.org/abs/hep-th/0003117}{{\tt
  http://arXiv.org/abs/hep-th/0003117}}].

\bibitem{Gherghetta:2000qt}
T.~Gherghetta and A.~Pomarol, {\it Bulk fields and supersymmetry in a slice of
  {AdS}},  {\em Nucl. Phys.} {\bf B586} (2000) 141--162,
  [\href{http://xxx.lanl.gov/abs/http://arXiv.org/abs/hep-ph/0003129}{{\tt
  http://arXiv.org/abs/hep-ph/0003129}}].

\bibitem{SUSY-RS2}
T.~Gherghetta and A.~Pomarol, {\it A warped supersymmetric standard model},
  {\em Nucl. Phys.} {\bf B602} (2001) 3--22,
  [\href{http://xxx.lanl.gov/abs/http://arXiv.org/abs/hep-ph/0012378}{{\tt
  http://arXiv.org/abs/hep-ph/0012378}}].

\bibitem{SUSY-RS3}
W.~D. Goldberger, Y.~Nomura, and D.~R. Smith, {\it Warped supersymmetric grand
  unification},
  \href{http://xxx.lanl.gov/abs/http://arXiv.org/abs/hep-ph/0209158}{{\tt
  http://arXiv.org/abs/hep-ph/0209158}}.

\bibitem{allanach}
B.~C. Allanach, K.~Odagiri, M.~A. Parker, and B.~R. Webber, {\it Searching for
  narrow graviton resonances with the {ATLAS} detector at {the Large Hadron
  Collider}},  {\em JHEP} {\bf 09} (2000) 019,
  [\href{http://xxx.lanl.gov/abs/http://arXiv.org/abs/hep-ph/0006114}{{\tt
  http://arXiv.org/abs/hep-ph/0006114}}].

\bibitem{Traczyk:2002jh}
P.~Traczyk and G.~Wrochna, {\it Search for {Randall-Sundrum} graviton
  excitations in the {CMS} experiment},
  \href{http://xxx.lanl.gov/abs/http://arXiv.org/abs/hep-ex/0207061}{{\tt
  http://arXiv.org/abs/hep-ex/0207061}}.

\bibitem{Davoudiasl:1999tf}
H.~Davoudiasl, J.~L. Hewett, and T.~G. Rizzo, {\it Bulk gauge fields in the
  {Randall-Sundrum} model},  {\em Phys. Lett.} {\bf B473} (2000) 43--49,
  [\href{http://xxx.lanl.gov/abs/http://arXiv.org/abs/hep-ph/9911262}{{\tt
  http://arXiv.org/abs/hep-ph/9911262}}].

\bibitem{Davoudiasl:2000wi}
H.~Davoudiasl, J.~L. Hewett, and T.~G. Rizzo, {\it Experimental probes of
  localized gravity: On and off the wall},  {\em Phys. Rev.} {\bf D63} (2001)
  075004,
  [\href{http://xxx.lanl.gov/abs/http://arXiv.org/abs/hep-ph/0006041}{{\tt
  http://arXiv.org/abs/hep-ph/0006041}}].

\bibitem{Davoudiasl:1999jd}
H.~Davoudiasl, J.~L. Hewett, and T.~G. Rizzo, {\it Phenomenology of the
  {Randall-Sundrum} gauge hierarchy model},  {\em Phys. Rev. Lett.} {\bf 84}
  (2000) 2080,
  [\href{http://xxx.lanl.gov/abs/http://arXiv.org/abs/hep-ph/9909255}{{\tt
  http://arXiv.org/abs/hep-ph/9909255}}].

\bibitem{Arkani-Hamed:1998rs}
N.~Arkani-Hamed, S.~Dimopoulos, and G.~R. Dvali, {\it The hierarchy problem and
  new dimensions at a millimeter},  {\em Phys. Lett.} {\bf B429} (1998)
  263--272,
  [\href{http://xxx.lanl.gov/abs/http://arXiv.org/abs/hep-ph/9803315}{{\tt
  http://arXiv.org/abs/hep-ph/9803315}}].

\bibitem{Antoniadis:1998ig}
I.~Antoniadis, N.~Arkani-Hamed, S.~Dimopoulos, and G.~R. Dvali, {\it New
  dimensions at a millimeter to a fermi and superstrings at a tev},  {\em Phys.
  Lett.} {\bf B436} (1998) 257--263,
  [\href{http://xxx.lanl.gov/abs/http://arXiv.org/abs/hep-ph/9804398}{{\tt
  http://arXiv.org/abs/hep-ph/9804398}}].

\bibitem{Accomando:1999sj}
E.~Accomando, I.~Antoniadis, and K.~Benakli, {\it Looking for {TeV}-scale
  strings and extra-dimensions},  {\em Nucl. Phys.} {\bf B579} (2000) 3--16,
  [\href{http://xxx.lanl.gov/abs/http://arXiv.org/abs/hep-ph/9912287}{{\tt
  http://arXiv.org/abs/hep-ph/9912287}}].

\bibitem{Antoniadis:1999bq}
I.~Antoniadis, K.~Benakli, and M.~Quiros, {\it Direct collider signatures of
  large extra dimensions},  {\em Phys. Lett.} {\bf B460} (1999) 176--183,
  [\href{http://xxx.lanl.gov/abs/http://arXiv.org/abs/hep-ph/9905311}{{\tt
  http://arXiv.org/abs/hep-ph/9905311}}].

\bibitem{Vacavant:2000wz}
L.~Vacavant and I.~Hinchliffe, {\it Model independent extra-dimension
  signatures with {ATLAS}},
  \href{http://xxx.lanl.gov/abs/http://arXiv.org/abs/hep-ex/0005033}{{\tt
  http://arXiv.org/abs/hep-ex/0005033}}.

\bibitem{HERWIG6}
G.~Corcella {\em et.~al.}, {\it {HERWIG} 6: An event generator for hadron
  emission reactions with interfering gluons (including supersymmetric
  processes)},  {\em JHEP} {\bf 01} (2001) 010,
  [\href{http://xxx.lanl.gov/abs/http://arXiv.org/abs/hep-ph/0011363}{{\tt
  http://arXiv.org/abs/hep-ph/0011363}}].

\bibitem{HERWIG6.4}
G.~Corcella {\em et.~al.}, {\it {HERWIG} 6.4 release note},
  \href{http://xxx.lanl.gov/abs/http://arXiv.org/abs/hep-ph/0201201}{{\tt
  http://arXiv.org/abs/hep-ph/0201201}}.

\bibitem{ATLFAST2}
E.~Richter-Was, D.~Froidevaux, and L.~Poggioli, {\it {ATLFAST} 2.0: a fast
  simulation package for {ATLAS}},  {\em ATLAS Internal Note} (1998)
  ATL--PHYS--98--131.

\bibitem{Giudice:1998ck}
G.~F. Giudice, R.~Rattazzi, and J.~D. Wells, {\it Quantum gravity and extra
  dimensions at high-energy colliders},  {\em Nucl. Phys.} {\bf B544} (1999)
  3--38,
  [\href{http://xxx.lanl.gov/abs/http://arXiv.org/abs/hep-ph/9811291}{{\tt
  http://arXiv.org/abs/hep-ph/9811291}}].

\bibitem{Han:1998sg}
T.~Han, J.~D. Lykken, and R.-J. Zhang, {\it On {Kaluza-Klein} states from large
  extra dimensions},  {\em Phys. Rev.} {\bf D59} (1999) 105006,
  [\href{http://xxx.lanl.gov/abs/http://arXiv.org/abs/hep-ph/9811350}{{\tt
  http://arXiv.org/abs/hep-ph/9811350}}].

\bibitem{Owens:1991ej}
J.~F. Owens, {\it An updated set of parton distribution parametrizations},
  {\em Phys. Lett.} {\bf B266} (1991) 126--130.

\bibitem{Martin:1998np}
A.~D. Martin, R.~G. Roberts, W.~J. Stirling, and R.~S. Thorne, {\it Scheme
  dependence, leading order and higher twist studies of {MRST} partons},  {\em
  Phys. Lett.} {\bf B443} (1998) 301--307,
  [\href{http://xxx.lanl.gov/abs/http://arXiv.org/abs/hep-ph/9808371}{{\tt
  http://arXiv.org/abs/hep-ph/9808371}}].

\bibitem{Corcella:2001pi}
G.~Corcella {\em et.~al.}, {\it {HERWIG} 6.3 release note},
  \href{http://xxx.lanl.gov/abs/http://arXiv.org/abs/hep-ph/0107071}{{\tt
  http://arXiv.org/abs/hep-ph/0107071}}.

\bibitem{Thorne}
R.~S. Thorne, {\it Private communication}, .

\bibitem{ATLAS_TDR}
{\bf ATLAS} Collaboration, {\it Detector and physics performance technical
  design report},  {\em CERN/LHCC} (1999) 99--15.

\bibitem{Abe:1997gt}
{\bf CDF} Collaboration, F.~Abe {\em et.~al.}, {\it Limits on quark-lepton
  compositeness scales from dileptons produced  in 1.8-tev p anti-p
  collisions},  {\em Phys. Rev. Lett.} {\bf 79} (1997) 2198--2203.

\bibitem{Abbott:1998rr}
{\bf D0} Collaboration, B.~Abbott {\em et.~al.}, {\it Measurement of the
  high-mass drell-yan cross section and limits on  quark-electron compositeness
  scales},  {\em Phys. Rev. Lett.} {\bf 82} (1999) 4769--4774,
  [\href{http://xxx.lanl.gov/abs/http://arXiv.org/abs/hep-ex/9812010}{{\tt
  http://arXiv.org/abs/hep-ex/9812010}}].

\bibitem{TeVphotons}
{\bf CDF} Collaboration, F.~Abe {\em et.~al.}, {\it Measurement of the
  cross-section for production of two isolated prompt photons in anti-p p
  collisions at s**(1/2) = 1.8-tev},  {\em Phys. Rev. Lett.} {\bf 70} (1993)
  2232--2236.

\bibitem{ref:sridhar}
K.~Sridhar, {\it Constraining the randall-sundrum model using diphoton
  production at hadron colliders},  {\em JHEP} {\bf 05} (2001) 066,
  [\href{http://xxx.lanl.gov/abs/http://arXiv.org/abs/hep-ph/0103055}{{\tt
  http://arXiv.org/abs/hep-ph/0103055}}].

\bibitem{Bonciani:1998vc}
R.~Bonciani, S.~Catani, M.~L. Mangano, and P.~Nason, {\it {NLL} resummation of
  the heavy-quark hadroproduction cross-section},  {\em Nucl. Phys.} {\bf B529}
  (1998) 424--450,
  [\href{http://xxx.lanl.gov/abs/http://arXiv.org/abs/hep-ph/9801375}{{\tt
  http://arXiv.org/abs/hep-ph/9801375}}].

\bibitem{Giele:1990vh}
W.~T. Giele, T.~Matsuura, M.~H. Seymour, and B.~R. Webber, {\it W boson plus
  multijets at hadron colliders: {HERWIG} parton showers versus exact matrix
  elements}, . Contribution to Proc. of 1990 Summer Study on High Energy
  Physics: Research Directions for the Decade, Snowmass, CO, Jun 25 - Jul 13,
  1990.

\bibitem{Mulguisin}
I.~C. Park, {\it A new showering algorithm: Mulguisin},  {\em ATLAS Internal
  Communication} (1999) ATL--COM--PHYS--99--055.

\end{thebibliography}\endgroup

\end{document}